# ALD grown Zinc Oxide with controllable electrical properties

E. Guziewicz[1], M. Godlewski[1,2], L. Wachnicki[1], T.A. Krajewski[1], G. Luka[1],

S. Gieraltowska[1], R. Jakiela[1], A. Stonert[3], W. Lisowski[4], M. Krawczyk[4], J.W. Sobczak[4],

A. Jablonski[4]

[1]*Institute of Physics, Polish Academy of Science, Al. Lotników 32/46, 02-668 Warsaw, Poland*

[2] *Dept. Mathematics and Natural Sciences, College of Sciences UKSW, ul. Dewajtis 5, 01-815 Warsaw, Poland*

[3]*The Andrzej Soltan Institute for Nuclear Studies, ul. Hoża 69, 00-681 Warsaw, Poland*

[4] *Institute of Physical Chemistry, PAS, ul. Kasprzaka 44/52, 01-224 Warsaw, Poland*

**Abstract.**

The paper presents results for zinc oxide films grown at low temperature regime by Atomic Layer Deposition (ALD). We discuss electrical properties of such films and show that low temperature deposition results in oxygen-rich ZnO layers in which free carrier concentration is very low. For optimized ALD process it can reach the level of $10^{15}$ $cm^{-3}$, while mobility of electrons is between 20 and 50 $cm^2/V\cdot s$. Electrical parameters of ZnO films deposited by ALD at low temperature regime are appropriate for constructing of the ZnO-based p-n and Schottky junctions. We demonstrate that such junctions are characterized by the rectification ratio high enough to fulfill requirements of 3D memories and are deposited at temperature $100^oC$ which makes them appropriate for deposition on organic substrates.

## 1. Introduction

Starting from its invention in the mid-seventies Atomic Layer Deposition (ALD, previously known also as Atomic Layer Epitaxy, ALE) has been experiencing ever-increasing interest. Originally developed by Suntola [1] for the growth of polycrystalline and amorphous zinc sulphide thin films for electroluminescence displays, over the years it greatly extended the scope of its applications [2-3]. In the early eighties this growth method was regarded as a kind of technological curiosity and was applied to rather few materials of II-VI compounds like CdTe, (Cd,Mn)Te, and ZnS [4]. The main driving force for development of the technique was preparation of ultrathin semiconductor layers of precisely controlled thickness in the nanometer scale. By 1980 ALD was being used to make remarkably good large-area thin films electroluminescent (TFEL) displays based on zinc sulfide doped with manganese [4]. Probably the most famous ALD application in the field of EL displays was the large panel display that for fifteen years, starting from 1983, was operating at the Helsinki airport [5]. Since that time ALD began to attract considerable attention as a method for producing high quality thin films. In the late eighties such films became to be of great importance both scientifically for growing low-dimensional structures as well as technologically in integrated circuits and optoelectronic devices [6-9]. In the middle eighties a new class of semiconductor compounds, III-V materials like GaAs and InAs, were successfully grown by ALD. At that time about 100 publications based on ALD materials was annually published, and III-V compounds dominated the list of investigated materials. The number of publications based on semiconductor films grown by ALD increased gradually each year reaching almost 450 papers in 2000 and exceeding 1300 in 2010 [10].

There are a lot of methods that can be presently used for thin film deposition like CVD, sputtering, MOCVD, PLE, MBE and many others. However, there is difficult to point out even one except ALD that is able to combine thickness control of nanometer scale and effectiveness needed for industrial processing. In fact, despite of low growth rate at the vertical direction, the volume increase during ALD processes can be appreciable because of a large substrate diameter that in industrial processing can exceed even one meter size [11]. On the other hand, extremely precise processing and uniform deposition of a 2-3 nm thin dielectric film necessary for a metal-oxide-semiconductor field effect transistor (MOSFET) performed in a 45 nm node, has been successfully achieved by ALD. In 2007 for the first time Intel successfully applied the ALD technique to growth of $HfO_2$ as a gate dielectric in the highly miniaturized processor (Penryn, Model QX9650) [12]. Deposition of the dielectric film with a thickness of only 2-3 nm was a major challenge, that had not been met for a few previous years until such dielectric layers were deposited by ALD. Dielectric films grown by other methods did not retain a high dielectric constant of a bulk material.

Successful deposition of a very thin gate dielectric in highly integrated circuits makes ALD even a more attractive method for a semiconductor films growth, because it creates the opportunity to grow different parts of electronic devices like gate, channel and gate dielectrics using the same technology. The first examples of this approach have been already demonstrated [13-15].

In this work we show the possibilities that gives the ALD technique for zinc oxide, which currently is the most investigated compound semiconductor with more than 5000 papers appearing in 2011. This booming interest is related to a wide range of applications that cover such different fields as photovoltaics, optoelectronics, sensing, transparent organic electronics and many more. Recently ZnO is also intensively studied for as a material for a novel 3D electronics, especially for nonvolatile cross-bar memories and organic electronics, that both require low temperature processing [16-20]. ALD technology is beneficial in this field because it is a method that is predestinated to low temperature deposition as we will present in the next paragraph.

First ZnO process by ALD have been reported in the mid-eighties and was based on zinc acetate and water precursors [21]. Recently, processes with organometallic precursors like diethylzinc (DEZn) and dimethylzinc (DMZn) and water have been developed [22] and deposition temperature can be lowered below 200$^{o}$C. Since that time the number of papers on ZnO films grown by ALD increases every year and has exceeded 70 papers in 2011. The great interest in ZnO-ALD films is closely related to the low growth temperature, which is a unique feature of the ALD processes with organometallic precursors. In fact, ZnO films grown at low temperature regime (100$^{o}$C – 200$^{o}$C) are not dedicated to optoelectronic applications, which required epitaxial layers deposited at elevated temperature, but to series of novel electronic applications like transparent electronics in which deposition on plastic substrates plays an important role.

In the present paper we demonstrate that ZnO grown by ALD at low temperature regime reveals controllable electrical properties that might be successfully applied in many fields of novel electronics.

## 2. ALD processes at low temperature regime

Idea of the ALD technique is based on sequential and self-limiting chemical reactions which take place at the surface of the growing film [1-5]. In the ALD process reactants (called here precursors) are alternatively introduced into a growth chamber and precursor's doses are interrupted by purging the reaction chamber by inert gas. In Fig. 1 we schematically show four ALD cycles and pressure of each chemical component during the process. Because of time intervals between doses, precursors do not have any possibility to react in the volume of the chamber and can meet only the reacted species at the surface. Therefore in ALD processes very reactive precursors can be used and thus low deposition temperatures are possible. For this reason we can tell that ALD is a deposition method which is dedicated to low-temperature-growth.

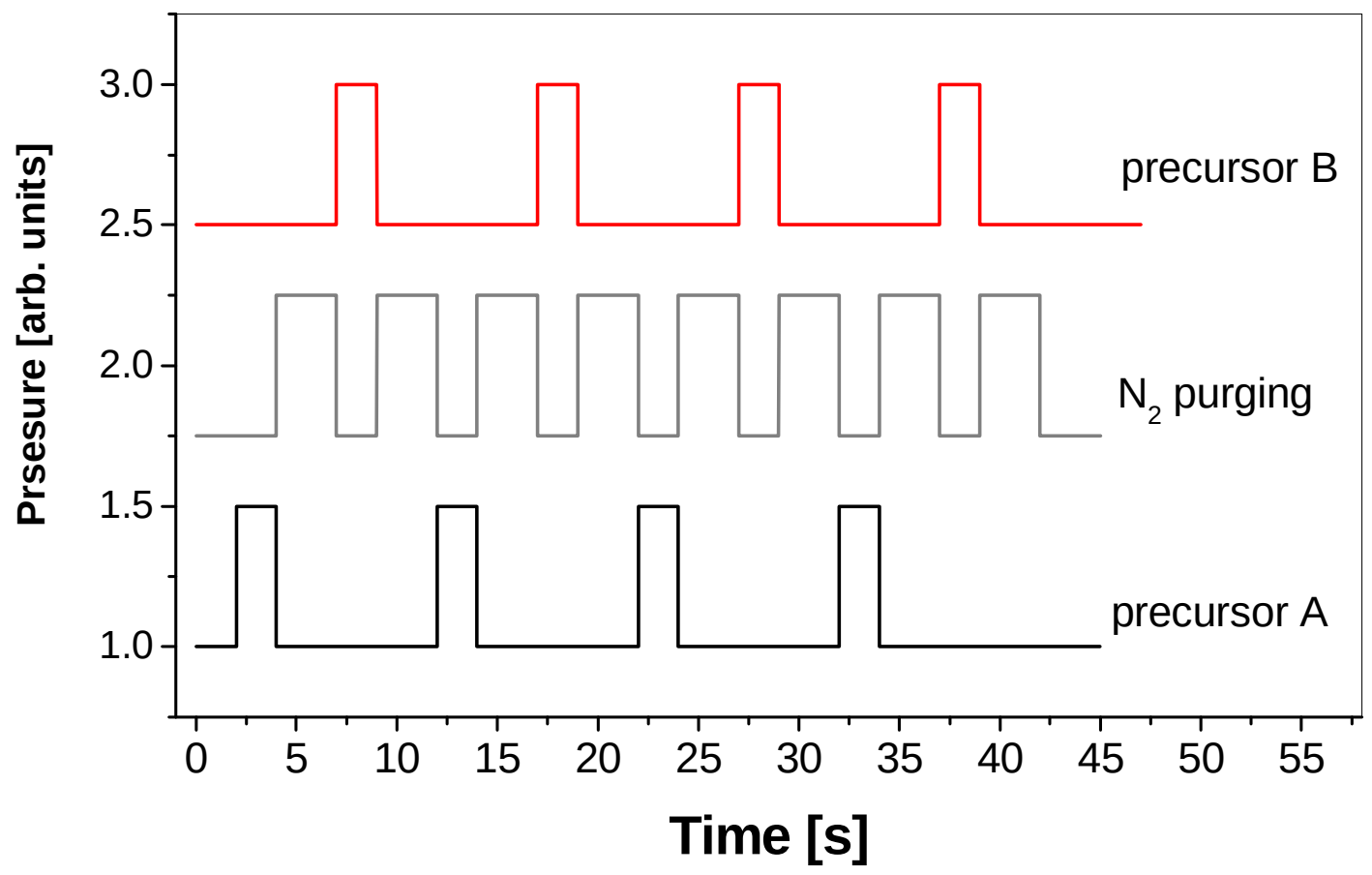

Fig. 1. The schematic view of the sequential procedure of the ALD process. Four ALD cycles are shown. For simplicity, both precursor's doses and purging times are shown the same, which is not usually the case in real processes.

In fact, the growth temperature used for deposition of a peculiar material depends on a kind of chemical reactants applied in the ALD process. Generally speaking, organic precursors have temperature and the difference usually exceeds 20$^{o}$C.

Zinc oxide in the ALD method can be obtained in several types of chemical reactions: synthesis from elemental precursors, single and double exchange. Because of high efficiency the last one is the most frequently used in ALD processes. Inorganic precursors (like Zn and O or $ZnCl_2$ and water) require growth temperature 400$^{o}$C – 500$^{o}$C and result in a rather low growth rate of about 0.5 Å per cycle [23-24]. The organic precursor zinc acetate, $Zn(CH_3COO)_2$, when used with water, enables decreasing deposition temperature to 300$^{o}$ – 380$^{o}$C with a comparably low growth rate.

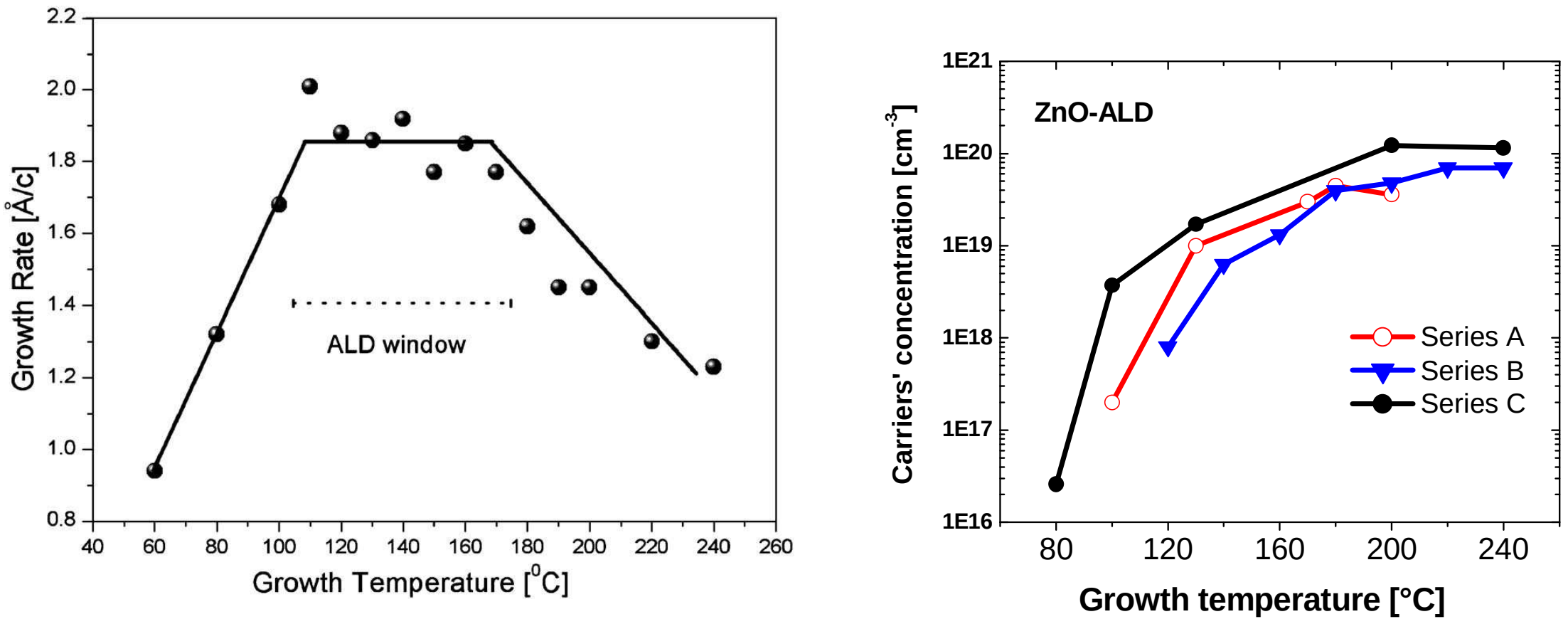

Fig. 2. (left) The ZnO growth rate versus growth temperature when DEZn and water are used as precursors for the ALD process. The ZnO films were 200 nm thick and deposited on Si substrate [after 25]. (right) Hall free carrier concentration versus growth temperature for series of 200 nm thick polycrystalline ZnO films deposited on a glass substrate and obtained by ALD with DEZn and water precursors [28].

Only organic precursors containing methyl and ethyl groups, like dimethylzinc (DMZn, $Zn(CH_3)_2$) or diethylzinc (DEZn, $Zn(C_2H_5)_2$), when used with deionized water, allow reducing growth temperature to 100$^o$C and below. Moreover, the ALD process efficiency in this case is much higher and inside the so-called "ALD growth window" exceeds 1.8 Å per cycle [Fig. 2 (left)].

The growth window is a term characteristic of ALD and related to the self-limited growth. It describes the temperature range where the balance between chemical reactivity and physical desorption is stable and thus a growth rate is constant with temperature [see Fig. 2 (left)]. The limits of the growth window are a characteristic feature of the used precursors and for DEZn and water they were established as 100$^o$C – 170$^o$C [26-28]. Although the ALD process can be performed outside of the growth window it is convenient to maintain it inside these temperature limits, because then thickness of the layer does not depend on temperature fluctuations and the film is perfectly uniform even at the atomic scale.

This does not means, however, that physical properties of layers grown at different temperatures from ‘the ALD window’ are the same. For example, free carrier concentration *n* of polycrystalline zinc oxide films scales with growth temperature even inside ‘the ALD window’ as shown in Fig. 2 (right). This means that when all parameters of the ALD process like precursor’s doses, purging times and pressures are the same, the electron concentration is lower in the ZnO film grown at lower temperature even inside the growth window. On the other hand, a change of growth parameters at the same growth temperature results in different free carrier concentration as can be seen for A, B and C series of samples in Fig. 2 (right). The A, B and C series of ALD processes differed with DEZn pulsing time, which was established as 0.02 s, 0.04 s and 0.06 s, respectively. The difference in carrier concentration for ZnO layers deposited at 100$^{o}$C is at the level of $10^{16}$ $cm^{-3}$ and is more than three orders of magnitude lower than *n* for ZnO films obtained with the same ALD parameters, but at temperature of 200$^{o}$C ($n \sim 10^{20}$ $cm^{-3}$).

The last finding is very important for applications, because it means that we can regulate electrical properties of ZnO films in wide limits, changing them from insulating to conductive, without any external doping, just by parameters of the ALD process. We thoroughly address this issue in the paragraph 5, where we show rectifying junctions based on ZnO-ALD films obtained at low temperature limits.

## 3. Structural and optical properties of ZnO-ALD films grown at low temperature regime

The results presented in this paper have been obtained for ZnO films grown in the Savannah-100 ALD reactor from Cambridge NanoTech with nitrogen used as a purging gas. ZnO films have been deposited on a silicon, silicon dioxide and glass substrates at growth temperature between 90$^{o}$C – 200$^{o}$C. Both precursors, diethylzinc and water, were kept at room temperature. The films grown within above temperature limits are polycrystalline as already reported [29]. We performed a few series of ALD processes with precursors' doses between 0.015 s and 0.06 s and purging time ranging from 8 s to 20 s. Before the growth process substrates were rinsed out in trichloroethylene, acetone and izopropanol and than cleaned in deionized water.

The preferred crystallographic orientation depends mainly on growth temperature and purging time. In order to achieve epitaxial growth ZnO films have to be deposited on GaN or $Al_2O_3$ templates and growth temperature should be $250^oC$ or higher [30].

Polycrystalline ZnO films obtained at low temperature regime are atomically flat and the Root Mean Square (RMS) value of the surface roughness depends on growth temperature and layer thickness and is between 0.5 and 4 nanometers [25, 29].

Polycrystalline ZnO films grown between $100^oC$ and $200^oC$ show both a band-edge and defect related photoluminescence [25, 31] although the former one is more intensive for films deposited at $130^oC$ and above. Anyway, the appearance of a blue band-edge emission is a fingerprint of good quality of obtained films.

## 4. Origin of electrical properties of ZnO-ALD films grown at low temperatures

The origin of scaling of electron concentration observed for ZnO-ALD films obtained at reduced growth temperatures is not clear. We expect that formation of defects like oxygen or zinc vacancies can be suppressed because it requires some activation energy which is not sufficient at low temperature regime.

The chemical reaction in the ALD process with DEZn and de-ionized water precursors can be express as follows:

$$C_2H_5 - Zn - C_2H_5 + H_2O \rightarrow ZnO + 2\ C_2H_6 \qquad (1)$$

This reaction occurs in two stage, when the half-part chemical reactions take place at the surface during DEZn (phase 1) and water (phase 2) pulses [32]:

DEZn phase: surface – OH + $C_2H_5$ – Zn – $C_2H_5$ $\rightarrow$ surface – O – Zn –$C_2H_5$ + $C_2H_6$ (2)

Water phase: surface – O – Zn –$C_2H_5$ + $H_2O$ $\rightarrow$ surface – O – Zn – OH + $C_2H_6$ (3)

Efficiencies of these two half-reactions change when temperature approaches to the water boiling point. In fact, for ZnO films grown at 100$^o$C, higher concentration of –OH groups has been found, which influences a film conductivity [33].

Conductivity of zinc oxide is a subject of a huge number of publications that have been appearing nowadays. Control of defects and associated free carrier concentration is of great importance for most of applications of this semiconducting compound. For nominally undoped ZnO a very high level of electron concentration *n* is commonly observed. Depending on the growth method and parameters used in the deposition process concentration between $10^{15}$ cm$^{-3}$ and $10^{20}$ cm$^{-3}$ has been reported [see 34 and references therein]. Based mainly on theoretical works [35-36] this very high level of doping has been attributed to zinc interstitials, oxygen vacancies and hydrogen which intrinsically involved in most of deposition processes. However, the problem has not been correctly addressed experimentally. In particular, the are only few experimental studies that relate the O:Zn atomic concentration in ZnO films with their electrical properties [37-40] and none of them have a systematic character.

In order to address this question we performed detailed XPS and RBS studies to determine atomic zinc and oxygen content in ZnO films grown at various temperatures. The problem is difficult to address, because we look for small differences in atomic composition of elements which total content in the material is high, close to 50%.

## a. XPS studies

We have performed the XPS studies on a series of polycrystalline ZnO films grown by ALD with DEZn and water precursors. The films were about 200 nm thick and deposited on silicon substrate at different temperatures while the rest of process parameters (precursors' doses, pressures and purging times) were the same. The commercial bulk ZnO crystal (MaTecK$^{TM}$) has been measured for comparison. Prior to XPS measurements the films were sputter-cleaned by 500V $Ar^{+}$ ions for 4 min. As a result, the XPS analysis was carried out on the ZnO films after removal of 6 nm of the material from the surface. The XPS spectra were recorded on the PHI 5000 VersaProbe$^{TM}$ scanning ESCA Microprobe using monochromatic Al-$K_{\alpha}$ radiation (hν = 1486.6 eV) from an X-ray source operating at 200 μm spot size, 50 W and 15 kV. The high-resolution XPS spectra were collected with the analyzer pass energy of 23.5 eV, the energy step 0.1 eV and the take-off angle 45$^{o}$ with respect to the surface plane.

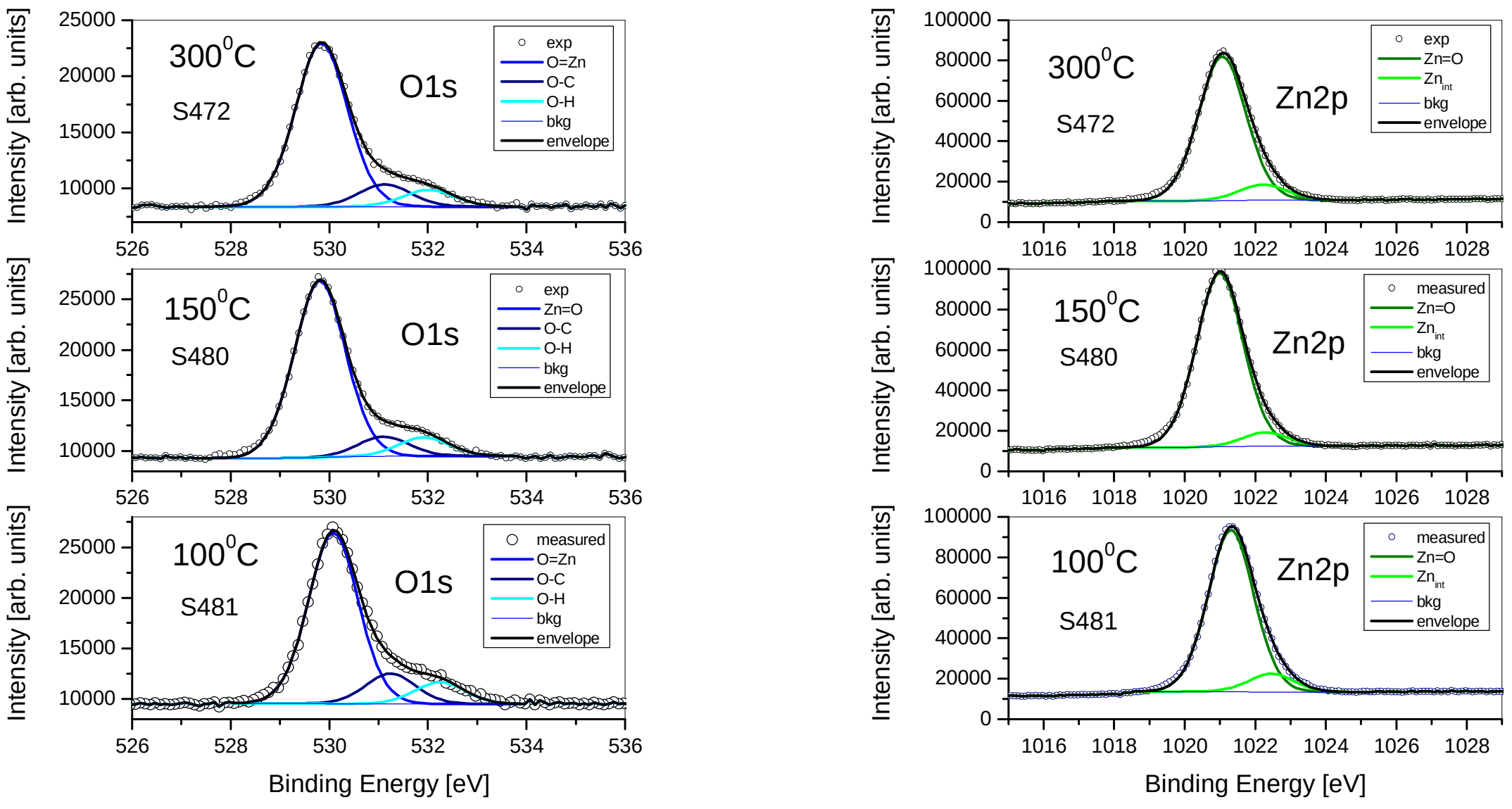

Fig. 3. (color online) O1s (left) and Zn2p (right) core level spectra measured for a series of polycrystalline ZnO films grown at 100$^{o}$C (bottom), 150$^{o}$C (middle) and 300$^{o}$C (top).

Shirley background has been subtracted from the measured XPS spectra. Binding energy (BE) was referenced to the C1s core level (BE = 284.6 eV. The atomic concentration (AC) of zinc and oxygen were evaluated using the Multiline program [41], where integrated area of XPS peaks associated with the Zn 2s, Zn $2p_{3/2}$, Zn $2p_{1/2}$, Zn 3s, Zn 3p, Zn 3d and O1s core levels were used for calculation. The most intensive Zn 2p and O 1s XPS spectra recorded on the films grown at 100°C, 150°C and 300°C are presented in Fig. 3.

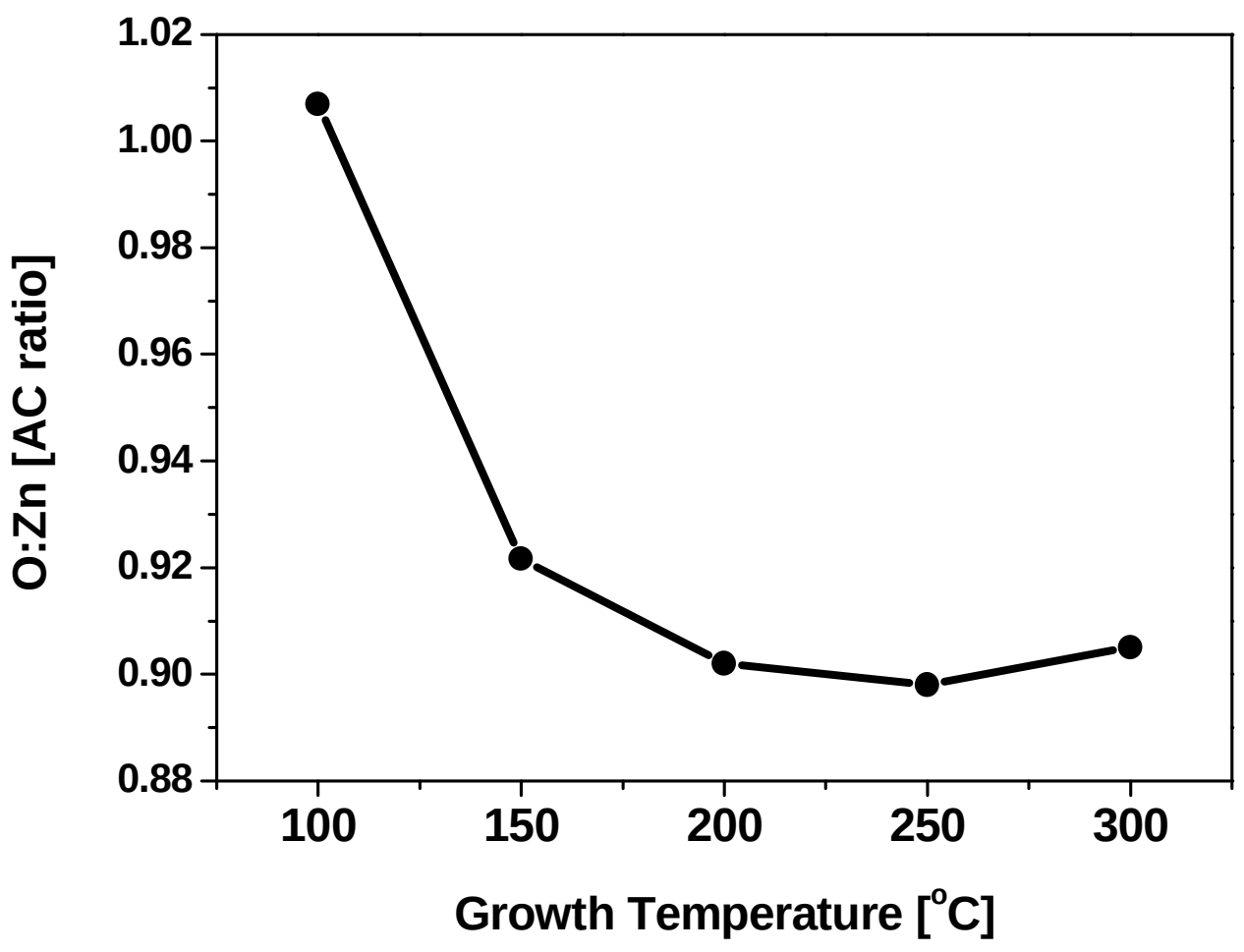

Fig. 4. The O/Zn atomic concentration ratio in ZnO-ALD films grown at temperatures 100-300 °C as derived from high-resolution XPS measurements. The O/Zn values are normalized to the O/Zn value (0.716) found on the commercial ZnO crystal (MaTecK™).

The shape of both XPS peaks is very similar, but we observe differences in the peaks intensity. In order to show how the temperature of film deposition influence the surface stoichiometry of the ZnO films, we compared the O/Zn AC ratio, evaluated for the films grown in various temperatures. Such a comparison is justified by application of the same argon ion sputter-cleaning procedure prior to the XPS analysis. We also assume that destructive effects induced by the argon etching are similar for all films. Values of the experimentally determined O:Zn ratio were normalized to the value obtained for the commercial ZnO single crystal (MaTecK™), which was also $Ar^+$ sputter- cleaned under the same experimental conditions. This calculation procedure was necessary to perform to get rid of influence of the sputtering process on the O:Zn ratio at the surface of the sample.

The sputtering is necessary to clean every surface before XPS measurements, because photoelectron spectroscopy is surface sensitive technique, so removal of impurities and residual gases have to be done before the experiment. On the other hand, when atomic masses of measured elements are very different, which is the case of Zn and O, the sputtering process influences the ratio between measured components as lighter elements are more effectively removed from the surface. In order to get rid of this artifact we compared the obtained values with the O:Zn value found for the commercial ZnO single crystal (MaTecK$^{TM}$). The results of normalized O:Zn values are illustrated in Fig. 4.

The O:Zn AC ratio takes the highest value for low growth temperature of ZnO films. This indicates oxygen content to be higher on samples grown at 100$^{o}$C and relatively lower at growth temperatures closer to 300$^{o}$C. As a result, the ZnO films prepared at the lowest temperature have the best Zn:O stoichiometry.

### b. RBS studies

A stoichiometry of zinc oxide layers was also analyzed with the RBS technique using 1.7 MeV 4He$^{+}$ ions from the Van de Graaff accelerator at the Helmholtz Zentrum Dresden-Rossendorf (HDZR), Germany. The experimental geometry was as follows: a scattering angle 160$^{o}$, a solid angle 3.135 mSr, energy resolution FWHM = 18 keV. All spectra were collected with the charge 40 µC and as rotating random in order to avoid the channeling effect. The SIMRA simulation code [42] was applied for spectra analysis.

We investigated a few series of polycrystalline zinc oxide films deposited on silicon which differed with ALD deposition temperature. In Fig. 5 (left) we show the typical RBS spectra of polycrystalline ZnO films.

RBS is a nondestructive technique which is used to quantitative depth profiling and aerial concentration measurements, especially for semiconductor thin films and multilayers. The method is based on elastic collisions between the light energetic projectile (usually $H^+$ or $He^+$ with energy up to several MeV) and target atoms. The energy loss of the ions gives depth information.

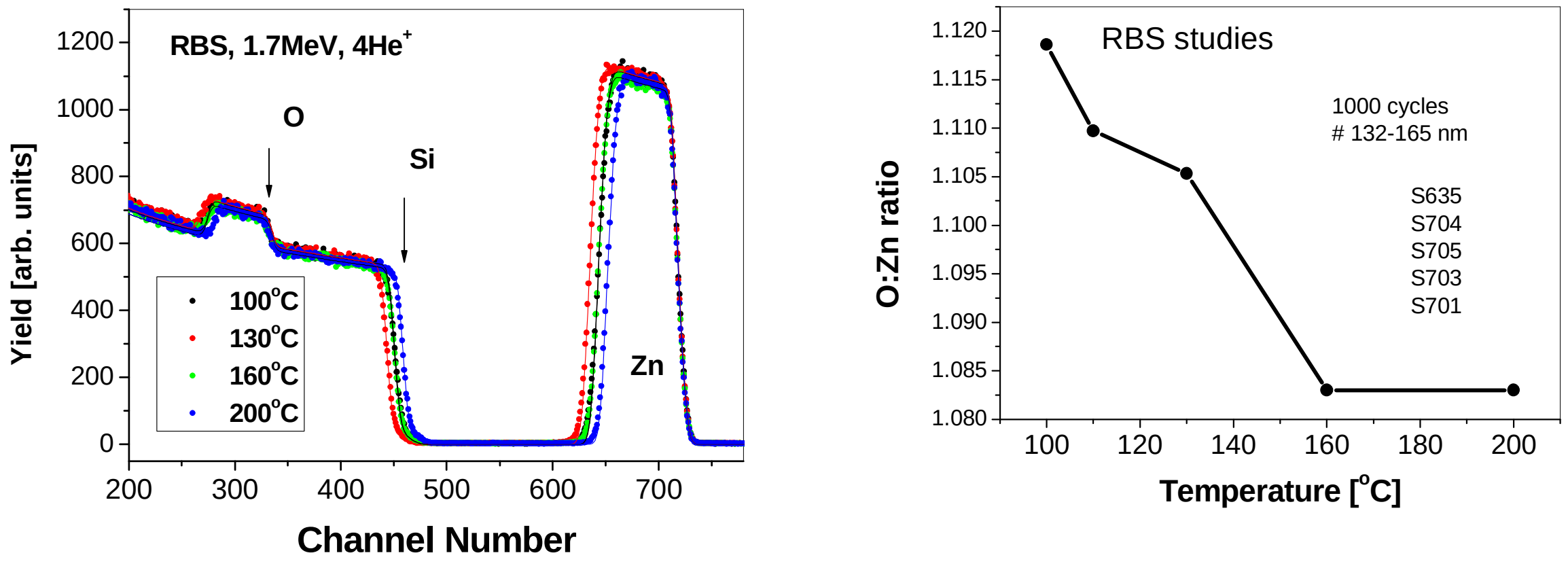

Fig. 5. Typical RBS spectra of a series of ZnO films grown by ALD at different temperatures (left), the O:Zn ratio for ZnO-ALD films calculated from the RBS results (right).

The film thickness can be calculated using the knowledge about the material density. Therefore the shape analyses of the emitted ions energy spectrum yields information about the concentration profiles.  However, RBS is not equally sensitive for every element. The RBS signal height depends on the scattering cross-section which is proportional to the square of atomic number of the element, so for the light elements a signal is much lower than for heavy ones [43]. Therefore it is quite difficult to separate the signal from such elements like nitrogen, carbon and oxygen, especially on heavier substrates.

Therefore from the RBS spectra we were able to evaluate the zinc content in every ZnO layer and we assumed that the remaining material is oxygen. We found that the O:Zn content calculated in this way varies between 1.08 to 0.99 depending on temperature and parameters of the ALD process. We also noticed that for every series of ZnO-ALD films the layers grown at low temperature have a higher oxygen content. The example of dependence between the O:Zn concentration and deposition temperature is shown in Fig. 5 (right). Temperature dependence is very clear and similar for every measured series. However, the exact numbers of the O:Zn content might be slightly different, because the analyses does not take into account the contant of light elements like nitrogen or carbon. This is also the origin of differences between XPS and RBS results, although in both cases the conclusion is the same: oxygen content is slightly higher in ZnO layers deposited at low temperature.

### c. Stoichiometry and electrical parameters of LT ZnO films

The XPS and RBS stoichiometry studies presented above contribute to understanding of the origin of free electron concentration in zinc oxide films grown at low temperature regime. In general, conductivity of zinc oxide crystals and films is strongly defect related and ranges from metallic to insulating depending of the deposition method and parameters used in the growth process. A reported free electron concentration in ZnO films varies usually from $10^{18}$ $cm^{-3}$ to $10^{19}$ $cm^{-3}$. The lower value of an electron concentration ($10^{15}$ – $10^{16}$ $cm^{-3}$) is typically achieved for K and Li doped bulk ZnO crystals grown by a hydrothermal method [44] or for ZnO films grown or annealed in oxygen at elevated temperature (700$^{o}$C and higher) [45]. Problem of high level of background electron doping in ZnO films and bulk crystals has been widely investigated, but, although progress is significant, there is still no definite answer, what causes high *n*-type conductivity in intentionally undoped ZnO. Native defects such as oxygen vacancy ($V_O$) or zinc interstitial ($Zn_i$) were indicated as defects responsible for high *n* carrier concentration in this material. However, theoretical works [46] supported by electron paramagnetic resonance experiments [47] indicate that oxygen vacancy ($V_O$) is a deep donor, while the formation energy of zinc interstitial ($Zn_i$) is relatively high, which points out that these defects cannot be responsible for a high *n* carrier concentration in zinc oxide [48].

Theoretical studies indicate on unintended introduction electrically active dopants during the growth process. It seems that hydrogen and carbon, which are present in many processes of crystal and thin films growth can be the dominant donors in this semiconducting material [49]. However, experimental data do not fully confirm this picture, because some experimental results indicate the presence of zinc interstitials in ZnO layers. The discrepancy between theoretical and experimental results emerges from the fact that theoretical studies relates to the equilibrium conditions, while the growth processes are generally carried out under conditions far from the equilibrium. Understanding the origin of background *n*-type conductivity of ZnO films would help control electrical properties of the material and thus would be important for future applications.

The XPS and RBS stoichiometry studies described in the previous paragraph show a clear correlation between a free electron concentration (see Fig. 2, left) and the O:Zn ratio (Fig. . These results point out that a low electron concentration observed for ZnO films grown at low temperature regime is related to a higher oxygen content in this material. The stoichiometry studies alone do not provide information on type of defects in the material, e.g. we cannot determine whether at low temperature conditions less oxygen vacancies or zinc interstitials are created. Low temperature photoluminescence studies [50], in which we observe a peak at 3.36 eV, indicate that rather the zinc interstitial is a dominant donor in ZnO films grown at low temperature, but we cannot exclude that formation of oxygen vacancies is also suppressed at low temperature.

The XPS and SIMS studies show that hydrogen and carbon impurities are not dominant donors in this material. The hydrogen content in the investigated ZnO layers does not scale with electron concentration, and it is even lower at elevated temperature (see Tab.1). This result suggests that hydrogen is not a dominant donor in the ZnO material, at least when the low temperature regime and the ALD process is concerned. The ToF-SIMS analysis [33] shows that ZnO layers deposited at 100$^o$C contain about twice more of OH groups than such layers deposited at 170$^o$C. The mechanism that causes trapping of OH groups at lower growth temperature may be based on an incomplete double exchange reaction between terminal OH groups and diethylzinc. Also at low temperature regime a small part of water molecules introduced as oxygen precursor can be trapped during the growth process. We suppose that the residual OH entities efficiently suppress formation of oxygen vacancies and in this way result in a low density of carriers.

| Growth T | 100$^o$C | 110$^o$C | 130$^o$C | 160$^o$C | 200$^o$C |
|---|---|---|---|---|---|
| Hydrogen content [%] | 0.58 | 0.46 | 0.38 | 0.34 | 0.33 |

Tab. 1. The hydrogen content in 400 nm thick ZnO layers deposited on a silicon substrate. The measurements were done by Secondary Ion Mass Spectroscopy (SIMS).

The carbon content in the ZnO films varies between 0.3 % to 2.1 % (see Tab.2) and is not correlated with electron concentration as shown in Tab. 2. The electron dispersive X-ray (EDX) analysis [51] leads to similar results.

Conclusion from the extensive chemical analysis presented above is as follows. Neither hydrogen nor carbon contaminations are dominant donors in ZnO films grown by ALD at low temperature regime. Low temperature deposition ensure conditions for the more stoichiometric growth, therefore zinc interstitials and oxygen vacancies formation is suppressed, which result in ZnO films with low carrier concentration. Such films are a very good material for electronic devices as will be shown in the next paragraph.

| Growth temp.[$^o$C] | 100$^o$C | 120$^o$C | 130$^o$C | 150$^o$C | 240$^o$C |
|---|---|---|---|---|---|
| Electron concentration [cm$^{-3}$] | $4\cdot 10^{16}$ | $1.5\cdot 10^{17}$ | $4\cdot 10^{18}$ | $3\cdot 10^{19}$ | $1\cdot 10^{20}$ |
| Carbon content | 1.3 | 2.1 | 0.3 | 0.8 | 0.4 |

Tab. 2. The carbon content (measured by XPS) and the electron concentration in 190 nm thick ZnO films grown on a silicon substrate.

## 5. Electronic applications

The correlation between free electron concentration and growth temperature observed for zinc oxide films grown by ALD with DEZn and water precursors is of great importance from the application point of view. This relation means that we can control electron concentration of ZnO films without any external doping but only by choosing appropriate parameters of the ALD process. In fact, different types of applications require various electrical parameters of the ZnO films. Theoretical calculations show [52] that electron concentration of zinc oxide film dedicated to Schottky junction should not exceed $2\cdot10^{17}$ $cm^{-3}$ while electron mobility should not be lower than 10 $cm^2/V\cdot s$. When ZnO is dedicated to a *p-n* junction, electron concentration at the level of $10^{18}$ $cm^{-3}$ is needed, and for ZnO films used as a transparent electrode for solar cells a very high $n$ , even at the level of $10^{21}$ $cm^{-3}$ is required.

Below we present examples that zinc oxide obtained at temperature of 100$^o$C by ALD can really work as an active element of rectification junctions, so that this material can be successfully applied in microelectronics.

Rectification junctions based on ZnO deposited at low temperature are elements that can be used as selectors in highly integrated non-volatile memories based on 3D cross-bar architecture [15-17, 33, 53-54]. Low temperature processing has to be implemented for both selector and storage elements in order to allow integration the Back-End-Of-the Line (BEOL) [20].

### a. Schottky diode

The selector element in the 3D cross-bar memory needs to fulfil specific requirements such as a high rectification ratio and a high forward current. The rectification ratio should be high enough to ensure proper selection of the appropriate memory cell without disturbance due to leakage current flowing through the remaining memory cells. A high forward current density is necessary to perform various operations (reading and writing) within the storage element. According to theoretical calculations [52] electron concentration in ZnO dedicated for a Schottky junction should not exceed $10^{17}$cm$^{-3}$, so taking into account temperature dependencies described in the paragraph 2 and shown in Fig. 2 (right), ZnO-ALD film dedicated for a Schottky junction was grown at temperature 100$^{o}$C.

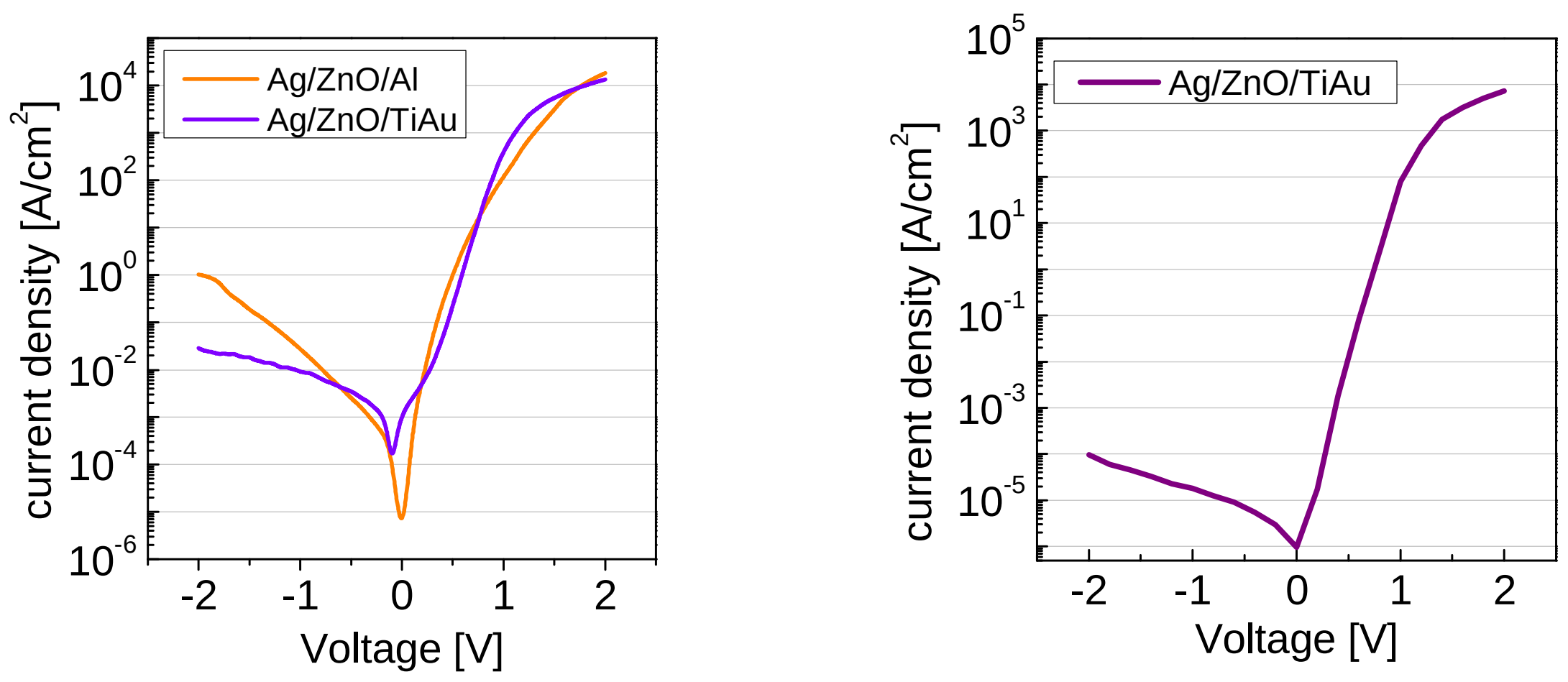

Fig. 6. The I-V characteristics of the Ag/ZnO-ALD Schottky junction obtained at 100$^{o}$C: contact optimization (left) and optimization of ALD growth parameters (right).

In the ALD process we used DEZn and deionized water precursors and we obtained a polycrystalline ZnO film with an intrinsic free carrier concentration $n = 10^{17}$cm$^{-3}$ and mobility of 17 cm$^2$/V·s.

A few possibilities of Schottky and ohmic contacts have been tested. Pt, Au and Ag have been checked as a Schottky contact and Al and Ti/Au as an ohmic contact. The choice of appropriate metal for a Schottky barrier is an important issue in a ZnO based Schottky junction, because the barrier height at the ZnO/metal interface does not follow the work function values [55]. This is because of ZnO surface states that strongly modify the barrier height. The best results we obtained for Ag as a Schottky and Ti/Au as an ohmic contact. Aluminium does not work properly as an ohmic contact, because it leads to a high leakage current as it is shown in Fig. 6 (left). Al is a known donor dopant in zinc oxide [56], so probably Al diffuses from the contact into a ZnO layer leading to a higher electron concentration and thus to a high leakage current due to increased tunnelling through a thinner barrier. The change of an ohmic contact from Al to Ti/Au leads to decrease of a leakage current by almost two orders of magnitude and subsequent increase of the junction rectification ratio from $3\cdot 10^3$ at 2V to $10^5$ (Fig. 6, left). The oxidation of silver at the Ag/ZnO interface causes raised series resistance. In order to minimize this effect a rectification Ag contact was deposited at the bottom of a structure and an ohmic contact at the top. Both I-V characteristics presented in Fig. 6. are obtained for such a structure. Details of the contact optimization has been described elsewhere [57-58].

Further improvement of the leakage current has been achieved by optimization of the ZnO ALD growth parameters.  As it is presented in Fig. 2 (right) for the same growth temperature various electron concentration can be obtained depending on the used ALD parameters like pulses and purging times. The optimization procedure is based on an unique approach that relies on the correlation of optical and electrical properties. For the ZnO deposition process we choose growth parameters that lead to ZnO film with very low defect-related photoluminescence [28].

Optimization of ZnO parameters results in a further decrease of a leakage current. In this way a rectification ratio at the level of $10^8$ at 2V has been achieved as is presented in Fig. 6 (right). Such a very high $I_{ON}/I_{OFF}$ value fulfils requirements for a switching element dedicated to a cross-bar memory. It assures the proper functioning of the crossbar array at very large integration scale. Moreover, the diode exhibit a high current density of $10^4$ A/cm$^2$. This Schottky junction properties are among the best results published so far for diodes obtained at low temperature regime [59]. The 10 kbyte crossbar memory array with a NiO based MIM (metal-insulator-metal) memory element and a Ag/ZnO-ALD/TiAu switching element was successfully constructed and showed appropriate switching properties [58]. It should be noted that the high rectification ration might be further increased by deposition of a thin 2-3 nm thick $HfO_2$ layer which positively influence the $I_{ON}/I_{OFF}$ ratio as was recently reported [60].

### b. p-type ZnO and ZnO-based homojunction

Obtaining p-type ZnO is a very challenging task, because this semiconductor compound is naturally n-type and self-compensation phenomena are likely to occur nullifying doping effect. To achieve p-type ZnO we explored several approaches, but the most successful procedure was the in-situ nitrogen doping during the ALD growth followed by annealing. Nitrogen was introduced into the ZnO film by changing the oxygen precursor from deionized water to ammonia water, while the rest of deposition parameters were the same. The samples were grown at 100$^o$C, pulsing and purging time were 0.015 s and 20 s for water (ammonia water) and 0.06 s and 8 s for the DEZn precursor. In this way concentration of nitrogen in the ZnO layer was significantly enhanced as it is shown in Fig. 7.

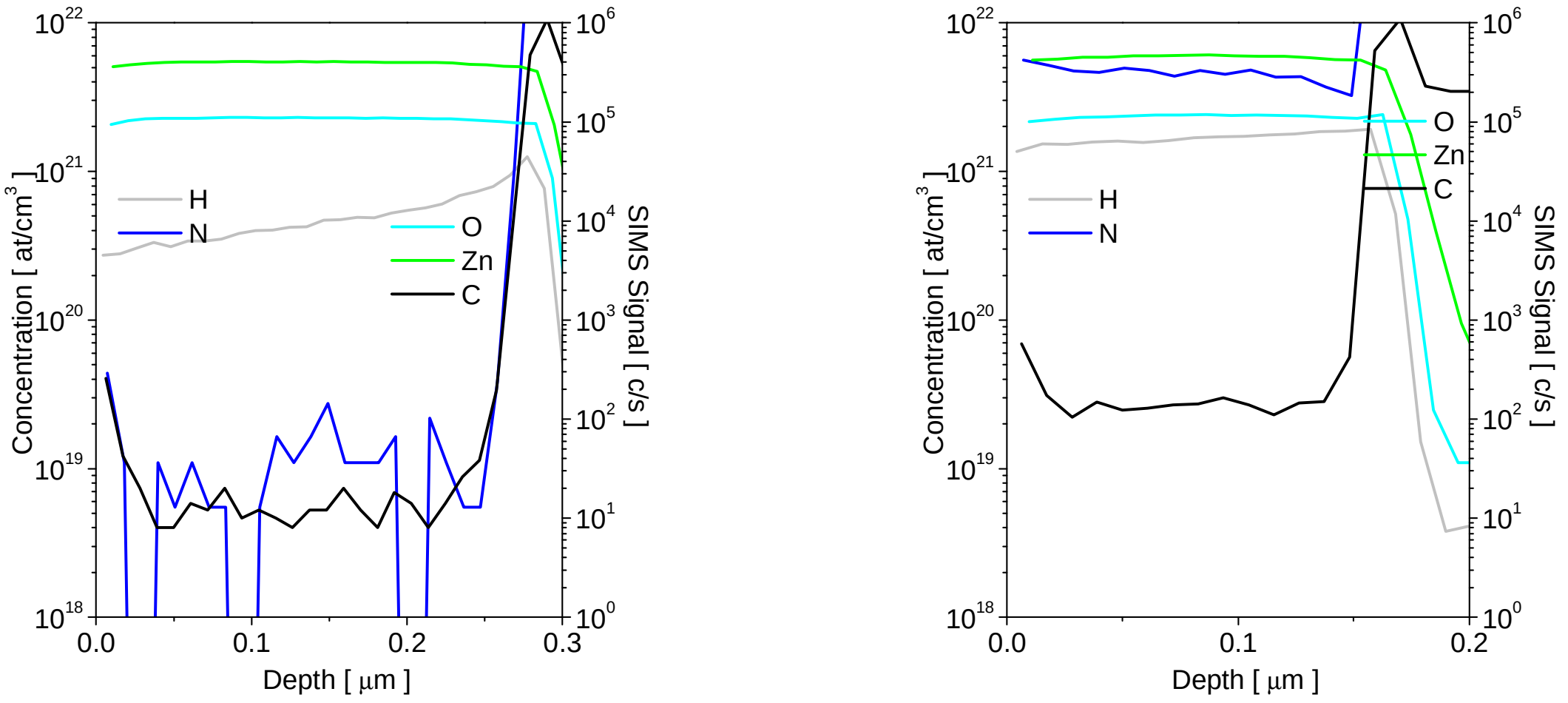

Fig. 7. SIMS depth profiles of undoped ZnO films (left) and ZnO film N-doped during the growth process (right).

After 4 minutes annealing at 350°C in nitrogen atmosphere ZnO films reveal p-type conductivity with a carrier concentration at the level of $10^{18}$ $cm^{-3}$ as was observed in the commercial Hall measurement system (RH2035 Phys. Tech GmbH). However, the obtained p-type ZnO films suffer from temporal instability. To avoid this effect the p-type ZnO layer was deposited as at the bottom of the junction i.e. the p-type ZnO layer was coated by n-type ZnO films, also grown in the low temperature ALD process.

With this procedure the ZnO based homojunction was obtained with a $I_{ON}/I_{OFF}$ ratio close to $10^5$ at 2V as it is presented in Fig. 8. The current density was $10^2$ $A/cm^2$. This parameters are not so perfect as for a Ag/ZnO-ALD Schottky junction described above.

However, these parameters are among the best parameters reported for ZnO based homojunctions published so far [61-63].

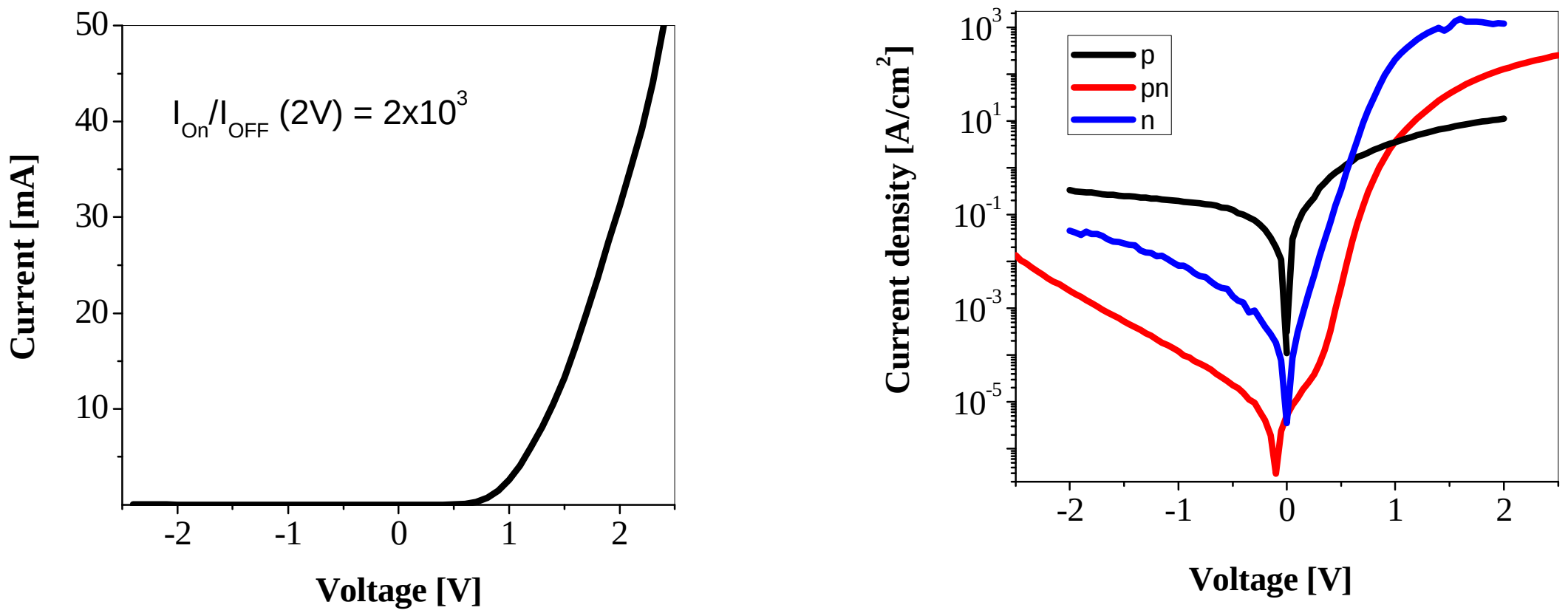

Fig. 8. ZnO homojunction obtained in the low temperature ALD process with both Ti/Au contacts (left) and Pt-p-ZnO and Au/Ti to n-ZnO contacts (right).

### c. Other electronic applications

II-VI semiconductors are nowadays studied as prospective candidates for thin film transistors (TFTs) dedicated for applications that require low processing temperature, such as transparent electronics or active matrix displays where the silicon based technology becomes less popular. Polycrystalline ZnO presents a few advantages over amorphous silicon such as much higher electron mobility and transparency. ALD is beneficial in this case, as not only ZnO films, but also the best quality dielectric layers, especially high-k oxides, can be obtained by this method. In that way the a few parts of field effect transistor, channel, gate dielectric and gate can be deposited using the same technology.

High quality films of high-dielectric oxides like $HfO_2$ and $Al_2O_3$ can be grown by ALD at temperature that does not exceed 100$^o$C [64]. The ALD method provides unique possibility to control thickness of the films at the nanometer scale. This is especially important for gate dielectrics used in MOSFET dedicated for highly integrated circuits where dielectric thickness is 3 – 5 nm. The average surface roughness of such thin dielectric films is at the level of 0.2 nm as it is shown in Fig. 9.

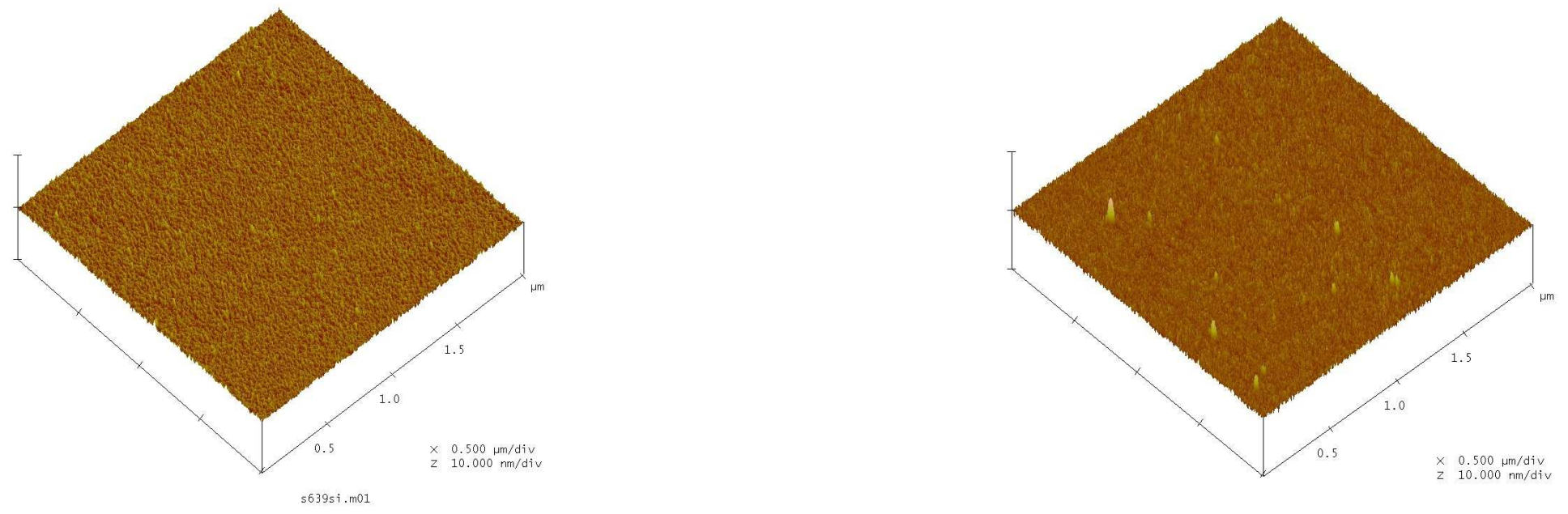

Fig. 9. AFM surface morphology of $HfO_2$ (left) and $Al_2O_3$ (right) 40 nm films deposited on silicon substrate. The RMS value of the surface roughness is 0.2 nm in both cases.

The MOSFET transistor with ZnO-ALD as a channel material and $Al_2O_3$ as a gate dielectric has been reported [33]. The ZnO layer deposited by ALD which acts as a channel there, was obtained at 100$^o$C and has carrier concentration below $10^{18}$ cm$^{-3}$. The device features a high $I_{ON}/I_{OFF}$ ratio of $10^7$.

Transparent thin film transistor (TTFT) where gate, gate dielectric and channel were obtained by ALD at low temperature regime was also reported [13]. Transistor structure was deposited on a glass substrate and two ZnO films with various conductivity were used as a gate and as a channel whereas a dielectric composite layer $Al_2O_2/HfO_2/Al_2O_3$ acts as a gate dielectric.

## 6. Summary

It has been shown that zinc oxide films deposited at a low temperature ALD process with DEZn and water precursors are naturally *n*-type with electron concentration that can be regulated without any external doping only by changing ALD process parameters. Achievable values of electron mobility and concentration are appropriate for applications in microelectronic devices like Schottky diodes and thin films transistors. *P*-type doping is also possible within the ALD process by replacing an oxygen precursor from deionized water to ammonia water. Nitrogen rich ZnO films obtained in this way show *p*-type conductivity after post-grown annealing. The exemplary application of such ZnO films has been presented. The rectification ratio of the ZnO-based homojunction is close to $10^5$ at 2V with the current density $10^2$ A/cm$^2$. The Schottky junction based on the Ag/ZnO-ALD structure reveals even better characteristic. Excellent parameters of the $10^8 = I_{ON}/I_{OFF}$ ratio and a current of $10^4$ A/cm$^2$ at 2V have been achieved, which are among the best results presented for diodes ever obtained for a polycrystalline material obtained at low temperature regime. These parameters fulfil requirements for the switching element dedicated for a 3D non-volatile memory.

The very important point is that all devices presented here have been obtained at very low temperature regime (100$^o$C) that is a crucial issue for many novel applications as 3D non-volatile memories and for temperature sensitive organic electronics. Moreover, zinc oxide films are transparent, therefore they can be used as active elements of transparent electronics. The ALD method has also been found as giving the best results for deposition of very thin high-k dielectric layers used as a gate dielectric in FET, so all the transistor structure (a gate, a gate dielectric and a channel) can be obtained using the only one deposition technique. The first results in this field were also mentioned.

To conclude, the ZnO grown at low temperature by the ALD technique is a prospective material for many novel electronic applications.

**Acknowledgements**

The research was partially supported by the European Union within European Regional Development Fund, through grant Innovative Economy (POIG.01.01.02-00-008/08) and by grants of the National Science Center of Poland (1669/B/H03/2011/40 and 2011/01/B/ST4/00959). The RBS measurements have been supported by the European Community as an Integrating Activity ‘Support of Public and Industrial Research Using Ion Beam Technology (SPIRIT)’ under EC contract no. 227012.

The authors acknowledge G. Tallarida and N. Huby, MDM, Milan, for contact deposition.

## References

1. Suntola T., Antson J., Method for producing Compound Thin Films, U.S. patent #4 058 430, issued Nov.25, 1977
2. Suntola T., in *Handbook of Crystal Growth*, vol. 3, part B: *Growth Mechanisms and Dynamics*; Hurle, D. T. J., Ed. By Elsevier, Amsterdam 1994, chapter 14
3. Suntola T., Thin Sol. Films **216**, 84-89 (1992)
4. Goodman, C. H. L., Pessa, M. V., J. Appl. Phys. **60**, R65 (1986)
5. George S.M., Chem. Rev. **110**, 111-131 (2010)
6. Niinisto L., Leskela M., Thin Sol. Films **225**, 130 (1993)
7. Ritala, M., Leskela M., in *Handbook of Thin Film Materials*, San Diego, USA (2001)
8. Leskela M., Ritala M., Thin Sol. Films **409**, 138 (2002)
9. Puurunen, R. L., J. Appl. Phys. **97**, 121301 (2005)
10. source : ISI Web of Knowledge
11. materials of 5th International Industrial Coating Seminar, 17-19 March 2010, Mikkeli, Finland, website : www.miics.net/archive/getfile.php?file=160
12. Ishimaru K., Solid-State Electron. **52**, 1266 (2008)
13. Gieraltowska S.A., Wachnicki L., Witkowski B.S., Godlewski M., Guziewicz E., Thin Solid Films, in print (doi:10.1016/j.tsf.2011.10.151)
14. Grundbacher R., Chikkadi K., Hierold C., J. Vac. Sc. Technol. **B28**, 1173 (2010)
15. Oh B.Y., Kim Y.H., Lee H.J., Kim B.Y., Park H.G., Han J.W., Heo G.S., Kim T.W., Kim JK.Y, and Seo D.S., Semicond. Sci. Technol. **26**, 085007 (2011)
16. Park B., Cho K., Kim S., Kim S., Nanoscale Res. Lett. **6**, 41 (2011)

17. Kasia E., Huby N., Tallarida G., Kutrzeba-Kotowska B., Perygo M., Ferrari S., Krebs F.C., Guziewicz E., Godlewski M., Osinniy V., Luka G., Appl. Phys. Lett. **94**, 143501 (2009)
18. Godlewski, E. Guziewicz, G. Łuka, T. Krajewski, M. Łukasiewicz, Ł. Wachnicki, A. Wachnicka, K. Kopalko, A. Sarem, B. Dalati, Thin Sol. Films (2009)
19. Luka G., Krajewski T., Wachnicki L., Szczepanik A., Fidelus J.D., Szczerbakow A., Lusakowska E., Kopalko K., Guziewicz E., Godlewski M., Acta Phys. Pol. A **114**, 1229 (2008)
20. Guziewicz E., Godlewski M., Krajewski T.A., Wachnicki Ł., Łuka G., Domagała J.Z., Paszkowicz W., Kowalski B.J., Witkowski B.S., Dużyńska A., Suchocki A., Phys. Stat. Sol. (b) **247**, 1611 (2010)
21. Tammenmaa M., Koskinen T., Hiltunen L., Niinistö L., Leakelä M., Thin Sol. Films **124**, 125 (1985)
22. Lujala V., Skarp J., Tammenmaa M., Suntola T., Appl. Surf. Sc. **82/83**, 34 (1994)
23. Butcher K.S.A., Afifuddin P., Chen P.T., Godlewski M., Szczerbakow A., Goldys E.M., Tansley T.L., Freitas J.A. Jr, J. Cryst. Growth **246**, 237 (2002)
24. Kopalko K., Godlewski M., Domagala J.Z., Lusakowska E., Minikayev R., Paszkowicz W., Szczerbakow A., Chem. Mat. **16**, 1447 (2004)
25. Guziewicz E, I.A. Kowalik, M. Godlewski, K. Kopalko, V. Osinniy, A. Wójcik, S. Yatsunenko, E. Łusakowska, W. Paszkowicz. J. Appl. Phys. **103**, 033515 (2008)
26. Makino H., Miyake A., Yamada T., Yamamoto N., Yamamoto T., Thin Sol. Films **517**, 3138 (2009)
27. Ku C.-S., Huang J.-M., Lin C.-M., Lee H.-Y., Thin Sol. Films **518**, 1373 (2009)
28. E. Guziewicz, M. Godlewski, T. Krajewski, Ł. Wachnicki, A. Szczepanik, K. Kopalko, A. Wójcik-Głodowska, E. Przeździecka, W. Paszkowicz, E.

Łusakowska, P. Kruszewski, N. Huby, G. Tallarida, and S. Ferrari, J. Appl. Phys. **105**, 122413 (2009)

29. Kowalik I.A., Guziewicz E., Kopalko K., Yatsunenko S., Wójcik-Głodowska A., Godlewski M., Dłużewski P., Łusakowska E., Paszkowicz W., J. Cryst. Growth **311**, 1096 (2009)
30. Wachnicki Ł., Krajewski T., Łuka G., Witkowski B., Kowalski B., Kopalko K., Domagala J.Z., Guziewicz M., Godlewski M, Guziewicz E., Thin Sol. Films **518**, 4556 (2010)
31. Guziewicz E., Godlewski M., Krajewski T.A., Wachnicki Ł., Łuka G., Paszkowicz W., Domagała J.Z., Przeździecka E., Łusakowska E., and Witkowski B.S., Acta Phys. Pol. A **116**, 814 (2009)
32. Ren J., Appl. Surf. Sc. **255**, 5742 (2009)
33. Huby N., Ferrari S., Guziewicz E., Godlewski M., Osinniy V., Appl. Phys. Lett. **92**, 023502 (2008)
34. Pearton S.J., Norton D.P., Ip K., Heo Y.W., and Steiner T., Superlattices Microstruct. **34**, 3 (2003).
35. Lany S., Zunger A., Phys. Rev. Lett. **98**, 045501 (2007)
36. Lany S., Zunger A., Phys. Rev. B**72**, 035215 (2005)
37. Kim H.S., Jung E.S., Lee W.J., Kim J.H., Ryu S.O., Choi S.Y., Ceramics Int. **34**, 1097 (2008)
38. Jeon S., Bang S., Lee S., Kwon S., Jeong W., Jeon H., Chang H.J., Park H.H., J. Electrochem. Soc. **155**, H738 (2008)
39. Ma Y., Du G.T., Yang T.P., Qiu D.L., Zhang X., Yang H.J., Zhang Y.T., Zhao B.J., Yang X.T., Liu D.L., J. Cryst. Growth **255**, 303 (2003)

40. Kwon S., Bang S., Lee S., Jeon S., Jeong W., Kim H., Gong S.C., Chang H.J., Park H.H., Jeon H., Semicond. Sci. Technol. **24**, 035015 (2009)
41. Jablonski A., Anal. Sci. **26**, 155-164 (2010)
42. Program for simulation of backscattering spectra for ion beam analysis with MeV, http://home.rzg.mpg.de/~mam/
43. Chu W.-K., Mayer J.W., Nicolet M.-A., *Backscatterin Spectrometry*, Academic Press, New York, San Francisco, London (1978)
44. Sekiguchi T., Miyashita S., Obara K., Shishido T., Sakagami N., J. Cryst. Growth **214-215**, 72, 2000
45. Lautenschlaeger S., Eisermann S., Hofmann M.N., Roemer U., Pinnisch M., Laufer A., Meyer B.K., von Wenckstern H., Lajn A., Schmidt F., Grundmann M., Blaesing J., Krost A., J. Cryst. Growth **312**, 2078 (2010)
46. Zhang S.B., Wie S.H., Zunger A., Phys. Rev. **B63**, 075205 (2001)
47. Kasai P., Phys. Rev. **130**, 989 (1963)
48. Kohan A.F., Ceder G., Morgan D., van de Walle C.G., Phys. Rev. **B61**, 15019 (2000)
49. van de Walle C.G., Phys. Rev. Lett. **85**, 1012 (2000)
50. Przeździecka E., Wachnicki Ł., Paszkowicz W., Łusakowska E., Krajewski T., Łuka G., Guziewicz E., Godlewski M., Semicon. Sci. Technol. **24**, 105014 (2009)
51. Krajewski T.A., Łuka G., Wachnicki Ł., Jakieła R., Witkowski B., Guziewicz E., Godlewski M., Huby N., Tallarida G., Optica Applicata **39**, 865 (2009)
52. Prà M., Schuster S., Erlen C., Csaba G., Lugli P., in: *Emerging Non-Volatile Memories, ESSDERC 2007*, München 2008

53. Lee M.J., Park Y., Suh D.S., Lee E.H., Seo S., Kim D.C., Jung R., Kang B.S., Ahn S.E., Lee C.B., Seo D.H., Cha Y.K., Yoo I.K., Kim J.S., Park B.H., Adv. Mat. **19**, 3919 (2007)
54. Lee M.J., Seo S., Kim D.C., Ahn S.E., Seo D.H., Yoo I.K., Baek I.G., Kim D.S., Byun I.S., Kim S.H., Hwang I.R., Kim J.S., Jeon S.H., Park B.H., Adv. Mat. **19**, 73 (2007)
55. Ip K., Thaler G.T., Yang H., Han S.Y., Li Y., Norton D.P., Pearton S.J., Jang S., Ren F., J. Cryst. Growth **287**, 149 (2006)
56. Fan J., Freer R., J. Appl. Phys. **77**, 4795 (1995)
57. Huby N., Tallarida G., Kutrzeba M., Ferrari S., Guziewicz E., Wachnicki Ł., Godlewski M., Microel. Eng. **85**, 2442 (2008)
58. Tallarida G., Huby N., Kutrzeba-Kotowska M., Spiga S., Arcari M., Csaba G., Lugli P., Radaelli A., Bez R., 2009 IEEE Int. Memory Workshop, Monterey, CA, May 10-14 2009, p. 6-9 (2009)
59. Allen M.W., Dubin S.M., Appl. Phys. Lett. **92**, 122110 (2008)
60. Krajewski T.A., Luka G., Gieraltowska S., Zakrzewski A.J., Smertenko P.S., Kruszewski P., Wachnicki L., Witkowski B.S., Lusakowska E., Jakiela R., Godlewski M., Guziewicz E., Appl. Phys. Lett. **98**, 263502 (2011)
61. Baltakesmez A., Tekmen S., Tüzemen S., J. Appl. Phys. **110**, 054502 (2011)
62. Gopalakrishnan N., Balakrishnan L., Pavai V.S., Elanchezhiyan J., Balasubramanian T., Curr. Appl. Phys. **11**, 834 (2011)
63. Tsai S.Y., Hon M.H., Lu Y.M., J. Cryst. Growth **326**, 85 (2011)
64. Gierałtowska S., Sztenkiel D., Guziewicz E., Godlewski M., Łuka G., Witkowski B.S., Wachnicki Ł., Łusakowska E., Dietl T., Sawicki M., Acta Phys. Pol. A **119**, 333 (2011)